\documentclass[journal]{IEEEtran}

\usepackage{xcolor,soul,framed} 

\colorlet{shadecolor}{yellow}
\usepackage[pdftex]{graphicx}
\graphicspath{{../pdf/}{../jpeg/}}
\DeclareGraphicsExtensions{.pdf,.jpeg,.png}

\usepackage[cmex10]{amsmath}
\usepackage{array}
\usepackage{mdwmath}
\usepackage{mdwtab}
\usepackage{eqparbox}
\usepackage{url}
\usepackage{hyperref}
 \usepackage{booktabs}
 \usepackage{soul}

\hypersetup{
    colorlinks=true,   
    linkcolor=blue,    
    urlcolor=black,     
    citecolor=blue,    
}

\bstctlcite{IEEE:BSTcontrol}

\begin{document}
\bstctlcite{IEEEexample:BSTcontrol}
    \title{Estimating Reliability of Electric Vehicle Charging Ecosystem using the Principle of Maximum Entropy}


\author{
}

\author{Himanshu Tripathi,
        Subash Neupane,
        Shahram Rahimi,
        Noorbakhsh Amiri Golilarz,
        Sudip Mittal,
        Mohammad Sepehrifar
\thanks{Himanshu Tripathi is with the Department of Computer Science and Engineering, Mississippi State University, MS, USA (e-mail: ht557@msstate.edu).}

\thanks{Subash Neupane is with the Department of Computer and Data Science, Meharry Medical College, Nashville, TN (e-mail: subash.neupane@mmc.edu).}

\thanks{Shahram Rahimi, Noorbakhsh Amiri Golilarz and Sudip Mittal are with the Department of Computer Science, The University of Alabama, Tuscaloosa, AL, USA (e-mail: srahimi1@ua.edu; noor.amiri@ua.edu; sudip.mittal@ua.edu).}

\thanks{Mohammad Sepehrifar is with the Department Of Mathematics and Statistics, Mississippi State University, MS, USA (e-mail: msepehrifar@math.msstate.edu).}
}

\markboth{}{Tripathi \MakeLowercase{\textit{et al.}}: Estimating Reliability of Electric Vehicle Charging Ecosystem using the Principle of Maximum Entropy}

\maketitle

\begin{abstract}
This paper addresses the critical challenge of estimating the reliability of an Electric Vehicle (EV) charging systems when facing risks such as overheating, unpredictable, weather, and cyberattacks. Traditional methods for predicting failures often rely on past data or limiting assumptions, making them ineffective for new or less common threats that results in failure. To solve this issue, we utilize the Principle of Maximum Entropy (PME) \cite{shannon1948mathematical, jaynes1982rationale}, a statistical tool that estimates risks even with limited information. PME works by balancing known constraints to create an unbiased predictions without guessing missing details. Using the EV charging ecosystem as a case study, we show how PME models stress factors responsible for failure. Our findings reveal a critical insight: even minor, localized stress events can trigger disproportionately large drops in overall system reliability, similar to a domino effect. The our PME model demonstrates how high-impact components, such as the power grid, are more likely to fail as stress accumulates, creating network-wide tipping points. Beyond EVs, this approach applies to any complex system with incomplete data, such as smart grids, healthcare devices, or logistics networks. By mathematically establishing an inverse relationship between uncertainty (entropy) and reliability, our work quantifies how greater system unpredictability directly degrades robustness. This offers a universal tool to improve decision-making under unpredictable conditions. This work bridges advanced mathematics with real-world engineering, providing actionable insights for policymakers and industries to build safer, more efficient systems in our increasingly connected world.
\end{abstract}


%
\IEEEpeerreviewmaketitle


\section{Introduction}
\begin{figure*}
    \centering
    \includegraphics[width=1\textwidth]{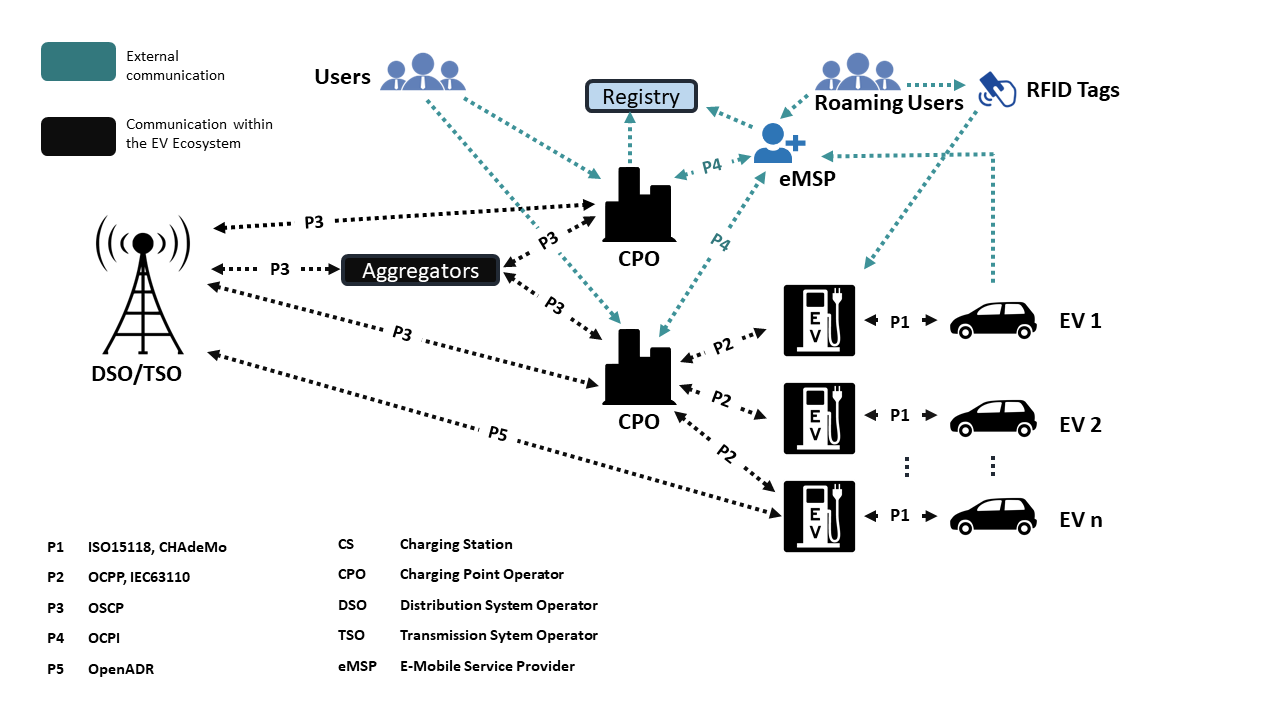}
    \caption{Illustration of the EV Charging Ecosystem: Communication Flow Between Grid, Charging Stations, Operators, and Users  \cite{kaur2025cybersecurity}}
    \label{fig:Ecosystem}
\end{figure*}

\IEEEPARstart{T}{he} widespread adoption of electric vehicles (EVs) hinges on the reliability of their charging ecosystems, which face different types of challenges ranging from cyberattacks to environmental stressors \cite{archana2020reliability}. These systems integrate power grids, charging stations, communication protocols, and user interfaces, all vulnerable to failures that cascade through the network. For instance, a compromised charging station software update can disable payment systems, while overheating cables during heatwaves accelerate hardware degradation \cite{ghavami2017reliability}. Such failures not only disrupt user experiences but also pose systemic risks, such as a single component malfunction, like a corrupted Radio-Frequency Identification (RFID) reader, that can render entire stations unusable, eroding public trust and increasing the amount of time that the network should take to perform its intended task \cite{pourmirza2021electric}. Combining these issues gives us a complex ecosystem, where interdependencies between utilities, manufacturers, and software providers create fragility. A software glitch in one subsystem might propagate unpredictably, delaying charging sessions and increasing stress on the affected components. Traditional reliability metrics, such as Mean Time Between Failures (MTBF) \cite{ghavami2017reliability}, struggle to quantify these dynamics because they rely on historical data ill-suited for new threats such as AI-driven cyber intrusions \cite{antoun2020detailed} or extreme weather events. As EV adoption surges, these reliability gaps threaten to undermine the green transition, necessitating methods that transcend conventional failure prediction frameworks.

While valuable in stable environments, existing reliability assessment tools often prove inadequate for the dynamic and unpredictable nature of EV networks. Many conventional techniques analyze system components in isolation, overlooking the complex, real-time interactions between hardware, software, and the electrical grid. For example, some models focus narrowly on software reliability without accounting for the physical degradation of hardware during rapid charging cycles, while others analyze the consumer feedback without integrating it with physical network parameters. Furthermore, cybersecurity assessments tend to prioritize threat mitigation over quantifying the impact of a security breach. In particular, they often overlook how such a breach (which has affected the amount of time that the network should have taken to perform its task concretely) alters failure probabilities across the network. A critical flaw shared by these approaches is their reliance on predictable failure modes and complete historical datasets, which makes them less effective against emergent, data-scarce scenarios such as zero-day exploits or unprecedented thermal stress factors. This limitation highlights the need for a paradigm shift toward a more holistic framework, one that can handle uncertainty and leverage minimal data to model risks without making assumptions.

PME emerges as a transformative solution in these cases, offering a mathematically rigorous way to estimate failure probabilities under uncertainty \cite{jaynes1982rationale}. Rooted in information theory, PME constructs the least biased probability distribution consistent with known constraints, such as average failure rates or temperature limits \cite{teitler1986maximum}. For instance, if engineers know a cooling fan of a charger fails twice annually but lacks humidity data, PME can estimate failure risks without assuming unverified patterns. This method maximizes uncertainty (entropy) within defined boundaries, ensuring predictions align with observable realities \cite{shannon1948mathematical}. Applied to EV charging, PME models factors such as power surges, component wear, hacking attempts and many more that can cause the system to take more time to charge a vehicle (which can be modelled as stress) as an input among with other variables, enabling dynamic risk assessments.

By correlating entropy with reliability, we can quantifies how unpredictability degrades system robustness. This adaptability makes PME uniquely suited for complex, evolving systems, where traditional methods oversimplify or overfit. Integrating PME with real-time data or digital twins \cite{neupane2023twinexplainer}\cite{neupane2022temporal} could revolutionize predictive maintenance, allowing operators to resolve failures by identifying stress accumulation before critical thresholds are reached.

This paper bridges advanced mathematics and practical engineering by formulating a PME-based reliability framework for EV charging ecosystems. The contribution of this work includes:
\begin{itemize}
    \item Formal representation of stress and failure probabilities.
    \item Formulating a closed-form analytical solution for modeling the reliability of a network (EV charging ecosystem).
    \item Analyzing the mathematical relationship between entropy and system/network reliability.
    \item Providing insights that can inform policy and infrastructure improvements.
\end{itemize} 

\noindent These contributions advance reliability engineering by addressing data scarcity and system complexity simultaneously. By validating the model through simulated case studies, the work demonstrates how PME is a fitting methods for predicting failures, especially the ones that are less common but catastrophic in nature \cite{tripathi2020maximum}. 

The remainder of this paper is organized as follows: Section II reviews the background and foundational concepts. Section III details the proposed methodology. Section IV presents a case study and its results. Section V discusses the broader applications of this framework, and Section VI concludes the paper with an outline for future work.

\section{Background and Foundational Concepts}
The operational integrity of the EV charging ecosystem relies on a complex, hierarchical flow of data and energy among multiple stakeholders as shown in Fig. \ref{fig:Ecosystem}, energy flows from Distribution System Operator (DSO) and Transmission System Operators (TSOs) to Charging Point Operators (CPOs), who manage station functionality via protocols like the Open Charge Point Protocol (OCPP) (Fig. \ref{fig:Ecosystem} P2). When users connects their vehicle, communication is established using standards like ISO 115118 (Fig. \ref{fig:Ecosystem} P1), while e-mobility service providers (eMSPs) handle authentication and billing for roaming users through protocols like the Open Charge Point Interface (OCPI) (Fig. \ref{fig:Ecosystem} P4). This intricate web of communication, while essential for functionality, introduces numerous vulnerabilities. For instance, load-altering attacks targeting aggregators can destabilize grid operations \cite{sanghvi2021cybersecurity}, and insecure physical ports can be exploited to install malware \cite{pourmirza2021electric}. The interconnectedness means that localized stress events, such as a station overload or a communication delay, can trigger cascading failures across the network, highlighting the challenges in modeling such a dynamic environment.

\begin{figure}
    \centering
    \includegraphics[width=1\linewidth]{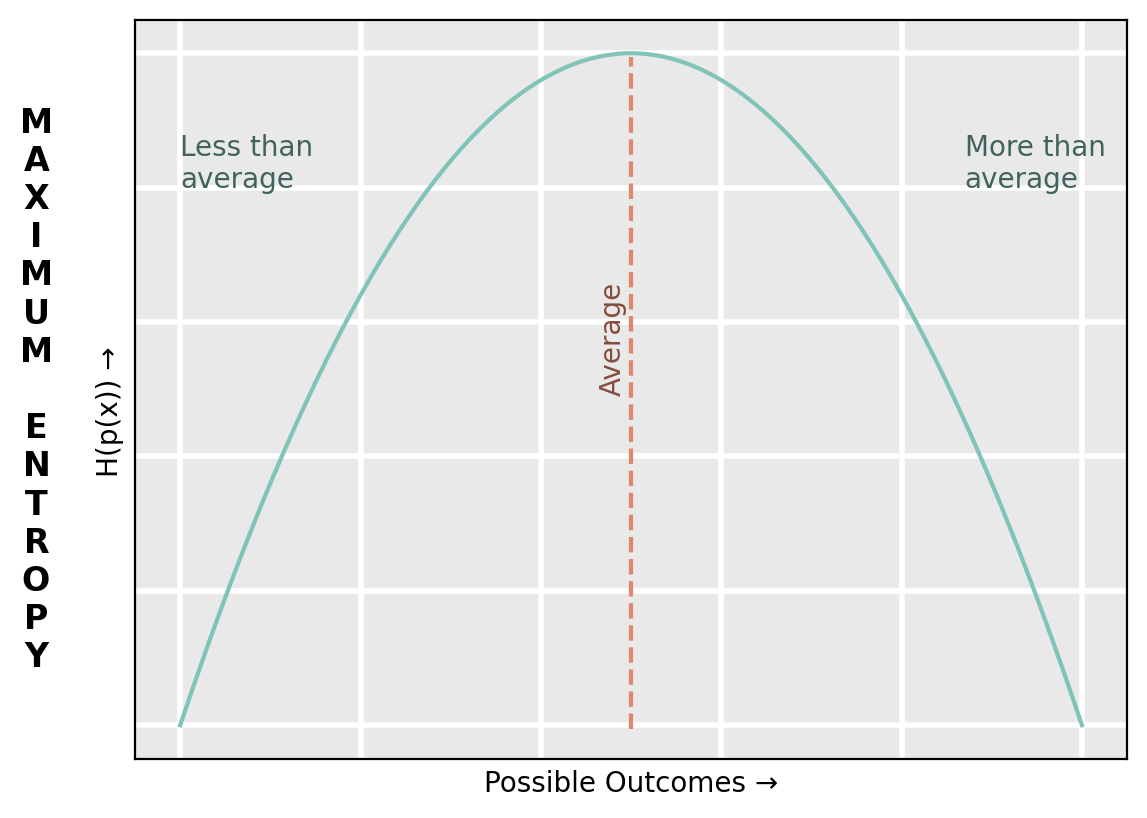}
    \caption{Entropy curve illustrating maximum uncertainty at the average outcome, with decreasing entropy as outcomes deviate from the average.}
    \label{fig:EntropyCurve}
\end{figure}

To analyze systems defined by such uncertainty and incomplete data, PME provides a robust mathematical foundation called Shannon's entropy, which was introduced by Claude E. Shannon in \cite{shannon1948mathematical} as a measure of informational uncertainty:

\begin{equation}
    \label{eq:Shannon}
    H(p(x)) = -\sum p(x) \cdot \ln(p(x))
\end{equation}

In Equation \ref{eq:Shannon}, $H$ is entropy $p(x)$ is the probability of an outcome $x$ from a random variable, and the natural log ($\ln$) simplifies combining probabilities, while the negative sign ensures entropy stays positive. PME derives the most unbiased probability distribution by maximizing Shannon's entropy (Equation \ref{eq:Shannon}), subject to known constraints, such as the total probability and known statistical moments of the system \cite{jaynes1982rationale}. When maximizing Shannon's entropy with respect to the constraints, the solution ends up as an exponential distribution. This distribution depends on Lagrange multipliers, which are set by the constraints applied to the system \cite{kang2009application}. This non-parametric approach and its concave nature (Fig. \ref{fig:EntropyCurve}) avoids unsupported assumptions about unobserved events and provides a unique solution, making it uniquely suitable for modeling the failure probabilities of components under novel or data-scarce stress conditions. The resulting Maximum Entropy Probability Distribution (MEPD) provides the foundational data needed for more advanced reliability assessments.

The distributions derived from PME are integral to modern reliability theory, which quantifies system durability through probabilistic metrics. A central concept is the hazard rate function, which represents the instantaneous rate of failure at time \textit{t} given that the system has survived until that point \cite{boland1997reliability}. For a system of components in series, the overall hazard rate is simply the sum of the individual component hazard rates, making it a powerful tool for comparative analysis, and the overall reliability of a system having components in series, which means if one component fails, the entire system fails, is given by: 

\begin{equation}
\label{eq:reliability}
    R(t) = \prod_{i=0}^n (1-\text{Failure Probability of the Components})
\end{equation}


where, \textbf{\textit{n}} is the total number of components and \textbf{\textit{R(t)}} is the reliability of the system, which is a function of time\footnote{\url{https://asq.org/quality-resources/reliability}}.


However, as Ghavami and Singh \cite{ghavami2017reliability} and Lu and Cai \cite{lu2017structural} note, judging reliability for systems like the EV charging ecosystem is not just about studying the behavior of each component individually but studying the entire system. Although studying about behavior of each component can help us to add backup parts \cite{boland1997reliability} which usually lowers failure chances, it does not always make the entire system more reliable or help us to determine when a component will fail. This shows that reliability analysis must study how the entire system works together, not just each component individually.

This need for a holistic perspective becomes evident when examining the current literature on EV charging reliability, which is largely fragmented. For instance, some frameworks rely on static grid-impact indices like SAIDI/SAIFI \cite{archana2020reliability}, which neglect real-time fluctuations in demand. Component-centric Markov models often overlook network-level interdependencies and the impact of cybersecurity threats \cite{ghavami2017reliability}. Meanwhile data-driven approaches analyzing consumer reviews can identify operational bottlenecks but lack integration with physical network parameters like line losses or voltage stability \cite{liu2023reliability}. Even scalable network reliability techniques are often too generalized, omitting EV-specific challenges such as the bidirectional power flow inherent in Vehicle-to-Grid (V2G) networks \cite{sahinoglu2010network}. This collective body of work, while valuable, leaves a critical gap: the absence of an integrated methodology that can simultaneously model real-time adaptability, multi-component interactions, and cyber-physical threats under data scarcity.

\section{Methodology}
\begin{figure}
    \centering
    \includegraphics[width=1\linewidth]{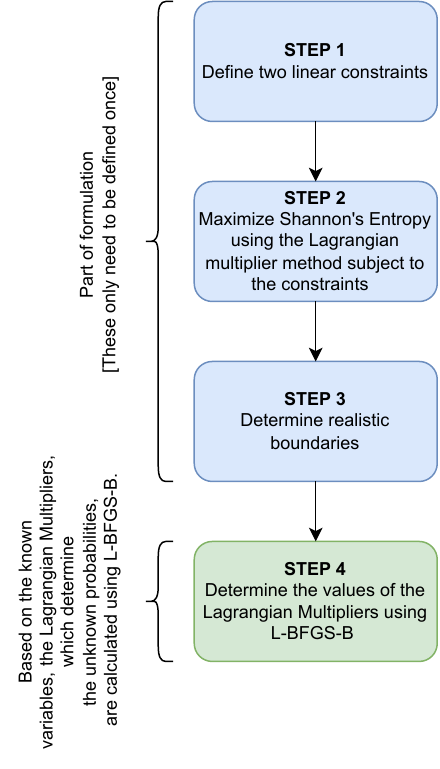}
    \caption{A four-step method for determining component failure probabilities: following the initial model formulation and boundary setting (Steps 1-3), only the final computational optimization (Step 4)  is needed for rapid recalculation with new network data.}
    \label{fig:steps}
\end{figure}

For our study, we have defined stress on any component as \textbf{the additional amount of time taken by the component to perform its intended task. In other words: if any EV is taking additional time to charge, that means at least one component must be taking more time to perform its intended task.}

This additional time can incur due to myriad stress factors including overheating, unpredictable weather, cyberattacks etc. In order to know how much time a car should have taken to charge we would need to calculate the required energy:

\begin{equation}
\label{eq:energy}
EN = \frac{BC}{100} \times (DSOC - ISOC)
\end{equation}





\noindent where, $EN$ refers to the energy needed in kWh, $BC$ denotes the battery capacity in kWh, $ISOC$ signifies the initial state of charge in percentage, and $DSOC$ represents the desired state of charge in percentage. Typically, the Desired State of Charge (DSOC) is the full capacity. The time required to charge the car is determined by:

\begin{equation}
\label{eq:timetaken}
    \text{Charging Time (hours)} = \frac{\text{EN}}{\text{CP} \times \text{CE}}
\end{equation}





\noindent where, $EN$ refers to the energy needed in kWh, $CP$ denotes the charging power in kW, and $CE$ represents the charging efficiency, which is generally 90\%, or 0.90.

The difference between the time taken by a car to charge and the expected amount of time that car should have taken to charge is our ``additional time". \textbf{To ensure clarity, we assume that only one component can fail or take additional time at any given time. This means that while multiple components can fail or incur additional time in sequence, no two components can do so simultaneously at the identical stress level (which is a rare scenario), even though the system is considered to have failed after the first component event.}

Let us assume an EV network comprises $``n"$ components and can go up to $``m"$ unit of additional time before a component is considered to have failed. Let $i^*$ denote the weakest component or the component that will fail at a given scenario. Therefore, the maximum operational stress level for the entire network, for that particular scenario will reach $m$ when $i^*$ will fail. Even if two components can fail simultaneously, the one which fails first will be $i^*$.

\begin{figure*}
    \centering
    \includegraphics[width=1\textwidth]{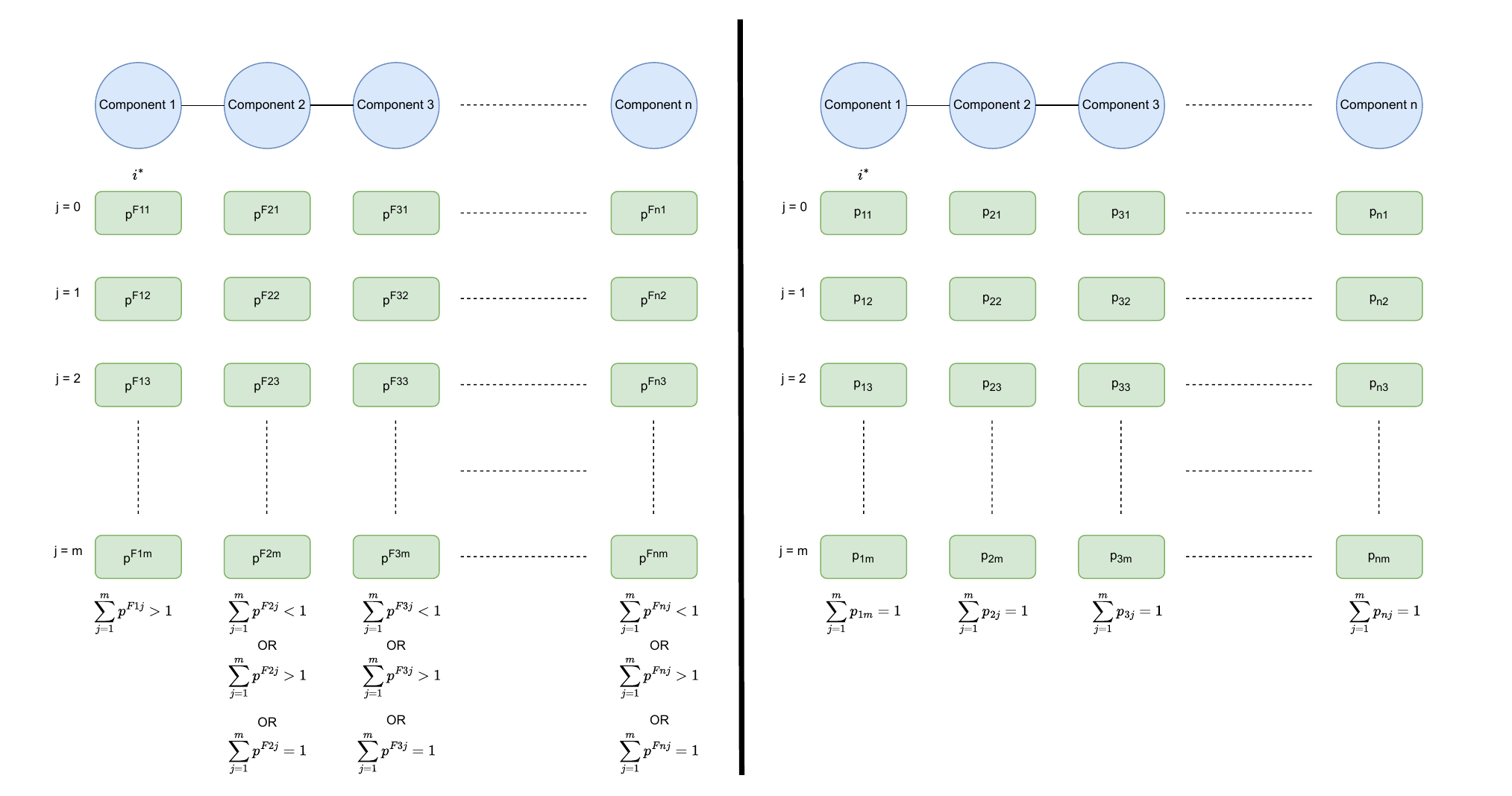}
    \caption{A contrast between the two foundational probability distributions for the components of the system. It visually separates the probability of a component experiencing a given stress level ($p_{ij}$) from the corresponding probability of failure ($p^{Fij}$), which is what the model ultimately seeks to determine.\\\textbf{Left:} Visual representation of the probability of failure, denoted as $p^{Fij}$, associated with each component state $i$. It graphically depicts key assumptions about how these failure probabilities behave across the stress range for this problem.\\\textbf{Right: }Visual representation of probability distribution, denoted as $p_{ij}$, for each component i facing the $j^{\text{th}}$ level of stress. The summations shown illustrate the fundamental property that, for any given component, the total probability of facing \textit{some} stress level must equal 1: $\sum_{j=1}^{m}p_{ij}=1$ for all i. The input probability values $p_{ij}$ used in subsequent calculations are assumed to satisfy this condition.}
    \label{fig:p_pF}
\end{figure*}

Since the network has a maximum stress limit, every additional charging delay has some failure chance. For component $i$ ($1 \le i \le n$) at stress level $j$ ($1 \le j \le m$), let $p^{Fij}$ be its failure probability. The total $p^{Fij}$ for component $i$ across all stress levels can be $> \text{ or } < \text{ or } = 1$, except for component $i^*$, where $\sum_i^m p^{Fij} > 1$ because failure probability grows as stress (additional charging time) increases, reaching 1 at $j=m$ (Fig. \ref{fig:p_pF} [Left]). Also, let $p_{ij}$ be the probability that the component $i$ causes $j$ units of additional charging time. For each $i$, all $p_{ij}$ values must add to 1 (Fig. \ref{fig:p_pF} [Right]). We assume $p_{ij}$ increases as $j$ increases. To model this system, we need to formally link component behaviors to network outcomes. This requires establishing two constraints that will form the basis for evaluating system-wide reliability. 

Our methodology for determining component failure probabilities follows a structured, four-step process (Fig. \ref{fig:steps}). This framework is partitioned into a one-time model formulation phase (Steps 1-3) and a repeatable computational phase (Step 4) for dynamic analysis. The formulation begins by defining two linear constraints that represent the total network failure probability and the total expected loss. Subject to these, we maximize Shannon’s entropy using the Lagrange multiplier method to derive an analytical expression for the failure probability ($p^{Fij}$) of any component at a given stress level. We then establish realistic boundaries for the Lagrange multipliers to ensure the subsequent optimization yields physically valid and stable solutions. This initial setup is performed only once. The final, operational step uses the Limited-memory Broyden–Fletcher–Goldfarb–Shanno algorithm with Bounds (L-BFGS-B) \cite{saputro2017limited} to compute the multiplier values based on new network data, allowing for rapid reliability recalculations. The progression through this framework, from initial formulation to dynamic computation, begins with establishing the foundational system constraints.

\subsection{Step 1: Defining Linear Constraints}
The first step in our modeling (Fig. \ref{fig:steps}) is to derive two linear constraints that can define the behavior of the network formally. The purpose of this formal representation and the related equations is to ensure that maximizing Shannon's entropy (Equation \ref{eq:Shannon}) under these constraints provides us an expression for the failure probability $p^{Fij}$. This entropy maximization approach provides an unbiased way to determine $p^{Fij}$, which is then used to calculate the system's reliability using Equation \ref{eq:reliability}.

\subsubsection{First Constraint: Network Failure Probability}
Let the total probability of failure of the entire network be $PF$. Unlike $p^{Fij}$, $PF$ is for the entire network, and not for any specific stress level $j$. 

\noindent It is given that the total probability of the failure can be evaluated as:

\begin{equation*}
    PF = 1 - P[\text{system survival}]
\end{equation*}

\noindent The entire system is considered failed if any single component fails:

\begin{align*}
PF = 1 - P[&\text{component 1 survives} \cap \text{component 2 survives} \\
&...\cap\text{component n survives}]
\end{align*}

\noindent We also know that the failure probability of any component is formulated as:

\begin{equation*}
    \text{Failure Probability of component \textit{i}} = \sum_{j=1}^{m} p_{ij}
    \cdot p^{Fij}
\end{equation*}

\noindent  where $0 \le \sum_{j=1}^m p^{Fij}\cdot p_{ij} \le 1$

\noindent Therefore, the survival probability of component \textit{i} is

\begin{equation*}
    P[\text{Component \textit{i} survives}] = 1 - \sum_{j=1}^{m} p_{ij} \cdot p^{Fij}
\end{equation*}

\noindent From the above equation the following equation can be driven as:

\begin{equation*}
    PF = 1 - \prod_{i=1}^n \left(1 - \sum_{j=1}^{m} p_{ij}
    \cdot p^{Fij}\right)
\end{equation*}

\noindent The following approximation was made: 

$\ln\left(1 - \sum_{j=1}^{m} p_{ij} \cdot p^{Fij}\right) \approx -\sum_{j=1}^{m} p_{ij} \cdot p^{Fij}$

\noindent Using the Taylor expansion of the products \cite{taylor1715methodus}:

\begin{align*}
\prod_{i=1}^{n}\left(1 - \sum_{j=1}^{m} p_{ij} \cdot p^{Fij}\right) \approx 1 &- \sum_{i=1}^{n} (\sum_{j=1}^{m} p_{ij} \cdot p^{Fij}) \\
&+ (\text{higher-order terms})
\end{align*}

\noindent Simplifying the summation:

\begin{align*}
\prod_{i=1}^{n} \left(1 - \sum_{j=1}^{m} p_{ij} \cdot p^{Fij}\right) \approx 1 &- \sum_{i=1}^{n} \sum_{j=1}^{m} p_{ij} \cdot p^{Fij} \\
&+ (\text{higher-order terms})
\end{align*}

\noindent Neglecting higher-order terms and minor simplification:

\begin{equation}
\label{eq:constraint1}
PF \approx \sum_{i=1}^{n} \sum_{j=1}^{m} p_{ij} \cdot p^{Fij}
\end{equation}

With this approximation, we have converted our equation from a multiplicative nature to a linear one, which is suited for being a constraint for entropy maximization. \textbf{Equation \ref{eq:constraint1} is our first constraint, representing PF, which is the total network failure probability, approximated as the sum of failure probabilities of each component weighted by its corresponding stress level likelihood.}

\subsubsection{Second Constraint: Expected Loss}
Let $UL_i$ ($UL_i \ge 1$) be \textbf{single unit loss for component $i$} (a fixed value, e.g., repair cost, downtime cost) and the expected total loss accumulated on all the components at $j^{th}$ stress level be $L_j$ as follows:

\begin{equation*}
    L_j =  \sum_{i = 1}^{n} UL_i \cdot p_{ij} \cdot p^{Fij}
\end{equation*}

\noindent Therefore, the total loss accumulated on all the components over all stress levels can be calculated as:

\begin{equation}
    \label{eq:constraint2}
    L = \sum_{j = 1}^{m}\sum_{i = 1}^{n} UL_i \cdot p_{ij} \cdot p^{Fij}
\end{equation}

\textbf{Equation} \ref{eq:constraint2} \textbf{is our second constraint, representing the total expected loss $L$, which is the sum of losses over all stress levels and all components, where the unit loss of each component is multiplied by its stress and failure probabilities.} Given two linear constraints (Equation \ref{eq:constraint1} and Equation \ref{eq:constraint2}), we maximize Shannon's entropy to find the most balanced probability distribution that satisfies these conditions. This approach minimizes unnecessary assumptions while fully using the available information.

\subsection{Step 2: Maximizing Entropy}

In the second step of our methodology (Fig. \ref{fig:steps}), we maximize Shannon's entropy (Equation \ref{eq:Shannon}). As mentioned earlier, we selected Shannon's entropy for its non-parametric nature. Unlike other entropy measures such as Rényi \cite{renyi1961measures} or Hartley entropy \cite{hartley1928transmission}, it requires no inputs beyond the known system constraints.


\noindent The objective is to maximize the entropy function:
\begin{equation*}
    H(p^{Fij}) = -\sum_{i = 1}^{n} \sum_{j = 1}^{m} p^{Fij} \ln (p^{Fij})
\end{equation*}
\noindent subject to the total network failure constraint (Equation \ref{eq:constraint1}):
\begin{equation*}
\sum_i^n \sum_j^m p_{ij} p^{Fij} - PF = 0
\end{equation*}
\noindent and the total expected loss constraint (Equation \ref{eq:constraint2}):
\begin{equation*}
\sum_i^n UL_i \sum_j^m p^{Fij} p_{ij} - L = 0
\end{equation*}

\noindent The constrained problem is solved by using the method of Lagrange multipliers \cite{lagrange1853mecanique}:

\begin{align*}
    \mathcal{L} = H(p^{Fij}) &- \lambda_1 (\sum_i^n \sum_j^m p_{ij} p^{Fij} - PF) \\&- \lambda_2 (\sum_i^n UL_i \sum_j^m p^{Fij} p_{ij} - L)
\end{align*}

\noindent Here, $\lambda_1$ and $\lambda_2$ are the Lagrange multipliers used to incorporate the constraints of the system. Specifically, $\lambda_1$ corresponds to the total network failure probability (PF), while $\lambda_2$ represents the total expected loss (L). Expanding the entropy term provides the full Lagrange function for maximization as follows:

\begin{align*}
     \mathcal{L} = -\sum_{i = 1}^{n} \sum_{j = 1}^{m} p^{Fij} \ln (p^{Fij}) &- \lambda_1 (\sum_i^n \sum_j^m p_{ij} p^{Fij} - PF) \\&- \lambda_2 (\sum_i^n UL_i \sum_j^m p^{Fij} p_{ij} - L)
\end{align*}

\noindent Partially differentiating this term with respect to $p^{Fij}$:

\begin{equation*}
    \frac{\partial \mathcal{L}}{\partial p^{Fij}} = -\ln(p^{Fij}) - 1 - \lambda_1(p_{ij}) - \lambda_2 (UL_i p_{ij})
\end{equation*}

\noindent In the above equation, setting $\frac{\partial \mathcal{L}}{\partial p^{Fij}} = 0$ identifies the maximum value of the function:

\begin{equation*}
    0 = -\ln(p^{Fij}) - 1 - \lambda_1(p_{ij}) - \lambda_2 (UL_i p_{ij})
\end{equation*}

\noindent Solving for the failure probability $p^{Fij}$,  yields an exponential form:

\begin{equation}
\label{eq:pF}
    p^{Fij} =  e^{- 1 - p_{ij}(\lambda_1 + \lambda_2 UL_i)}
\end{equation}

Equation \ref{eq:pF} is an expression for $p^{Fij}$ which we require to calculate for all the components and for all stress levels in order to determine the reliability of the entire system. This $p^{Fij}$ value is dependent on two Lagrange multipliers ($\lambda_1, \lambda_2$) as well as the probability of a component $i$ experiencing stress $j$ ($p_{ij}$) and unit loss for each component $i$ ($UL_i$). Additionally, the following condition must be considered to ensure the result is a valid probability:
\begin{equation}
\label{eq:pFlimit}
p_{ij}(\lambda_1 + \lambda_2 UL_i) > -1
\end{equation}

This is because value of $p^{Fij}$ cannot be more than 1, which means $e^0$ is the limit. This guarantees the exponent in the failure probability formula remains negative, ensuring valid probabilities. Before computing $p^{Fij}$ for each $i$ and $j$, we must define the feasible range for $\lambda_1$ and $\lambda_2$. This ensures the optimizer function (step 4, Fig. \ref{fig:steps}) yields valid values. Boundary conditions for Lagrange multipliers must be established prior to their computation phase.

\subsection{Step 3: Determining Boundary Values for Lagrange Multipliers}
Given the expression for $p^{Fij}$ in Equation \ref{eq:pF}, this step focuses on determining the boundary values for the two Lagrange multipliers (Fig. \ref{fig:steps}). Equation \ref{eq:pF} shows that $p^{Fij}$ is dependent on $p_{ij}$ and $UL_i$. Therefore, analyzing its behavior relative to these variables can define the boundary values for the Lagrange multipliers.

To analyze this relationship, the partial derivative of $ p^{Fij} $ with respect to $ p_{ij} $ is computed:
    \begin{equation*}
    \frac{\partial p^{Fij}}{\partial p_{ij}} = -(\lambda_1 + \lambda_2 UL_i) \cdot e^{-1 - p_{ij}(\lambda_1 + \lambda_2 UL_i)}
    \end{equation*}
\noindent Since the exponential term $ e^{-1 - p_{ij}(\lambda_1 + \lambda_2 UL_i)} $ is always positive, the sign of the derivative depends on $ -\left(\lambda_1 + \lambda_2 UL_i\right) $. Therefore, $\lambda_1 + \lambda_2 UL_i$ can have 3 possible cases:
    \begin{itemize}
        \item \textbf{Case 1}: If $ \lambda_1 + \lambda_2 UL_i > 0 $:
        \begin{equation*}
        \frac{\partial p^{Fij}}{\partial p_{ij}} < 0 \quad \Rightarrow \quad p^{Fij} \text{ decreases as } p_{ij} \text{ increases.}
        \end{equation*}
        
        \item \textbf{Case 2}: If $ \lambda_1 + \lambda_2 UL_i < 0 $:
        \begin{equation}
        \label{eq:lambdavalues}
        \frac{\partial p^{Fij}}{\partial p_{ij}} > 0 \quad \Rightarrow \quad p^{Fij} \text{ increases as } p_{ij} \text{ increases.}
        \end{equation}

        \item \textbf{Case 3}: If $ \lambda_1 + \lambda_2 UL_i = 0 $: $ p^{Fij} $ is constant (no dependence on $ p_{ij} $).
    \end{itemize}
    
Case 2 (Equation \ref{eq:lambdavalues}) represents the ideal scenario for our model. This is because as stress on any component increases, $p^{Fij}$ also increases. When this principle is combined with our assumption that the probability of a stress experienced by a component $i$ at some stress level $j$ ($p_{ij}$) also grows with the stress level, the model establishes a strong interdependence between $p^{Fij}$ and $p_{ij}$, that is given $p_{ij}$ increases with the stress level ($j$), it can be concluded that the value of $p^{Fij}$ must be increasing with $p_{ij}$. Essentially, this condition ensures that if a component has a high probability of experiencing a given stress level $p_{ij}$, its $p^{Fij}$ at that level is also high. This interdependence means that even a small increase in stress at critical points can cause a large jump in the $p^{Fij}$ (given the values for $UL_i, PF, L$ and the Langrangian multipliers), similar to a domino effect where minor disruptions lead to big failures. So, we investigate where the boundaries for $\lambda_1$ and $\lambda_2$ stand by considering the condition in Equation \ref{eq:lambdavalues} and Equation \ref{eq:pFlimit}.

Given Equation \ref{eq:lambdavalues}, there arises two possible cases:

\begin{itemize}
    \item $\lambda_1 > 0 \text{ and } \lambda_2 < 0$
    \item $\lambda_1 < 0 \text{ and } \lambda_2 > 0$
\end{itemize}

The Lagrange multipliers, $\lambda_1$ and $\lambda_2$, function as `shadow prices' that control the sensitivity of the uncertainty in the system. Each Lagrange multiplier is directly related to a system-wide value, $\lambda_1$ corresponds to the total network failure probability ($PF$) via Equation \ref{eq:constraint1}, while $\lambda_2$ corresponds to the total expected loss ($L$) via Equation \ref{eq:constraint2}. This highlights a limitation of the equations and its structure that is $\lambda_1$ and $\lambda_2$ cannot both be positive (Equation \ref{eq:lambdavalues}), which forces a design choice between one of the two cases shown above. \textbf{We can either opt for a positive $\lambda_1$ to show that as $PF$ increases system entropy also increases or a positive $\lambda_2$ to show that as overall loss $L$ increases entropy also increases, with the other lambda being forced into a negative value to satisfy the Equation \ref{eq:lambdavalues}.} Regardless of which foundational principle is selected, the core mathematics for the optimization would remain theoretically valid. The choice is based on the use case our model is applied on.

For our work, we opt for condition $\lambda_1 < 0$ and $\lambda_2 > 0$. This is because components with higher unit loss ($UL_i$), meaning those components that have higher repair costs or downtime, are typically well-engineered, meaning these components are less likely to fail. Therefore, as $UL_i$ increases $p^{Fij} \cdot p_{ij}$, must decrease. In mathematical terms, this inverse relationship means the rate of change (the partial derivative) of the risk term with respect to $UL_i$ must be negative:

\begin{equation*}
    \frac{\partial}{ \partial UL_i}p^{Fij} \cdot p_{ij} < 0
\end{equation*}

\noindent To analyze this condition further, we can substitute the formula for $p^{Fij}$ from Equation \ref{eq:pF}. This gives us the following inequality:

\begin{equation*}
p_{ij} \frac{\partial}{\partial UL_i} (e^{-1-p_{ij}(\lambda_1+\lambda_2 UL_i)}) < 0
\end{equation*}
\noindent Applying the chain rule for differentiation gives the following expanded expression:
\begin{equation*}
e^{-1-p_{ij}(\lambda_1+\lambda_2 UL_i)} \cdot (-p_{ij}\lambda_2) < 0
\end{equation*}

Since the exponential term is always positive and $p_{ij}$ values are probabilities, they are positive as well. Therefore, $\lambda_2>0$ so that our intution that higher $UL_i$ has lower $p^{Fij} \cdot p_{ij}$ holds. Given these conditions we need to compute the values for $\lambda_1, \lambda_2$.

\subsection{Step 4: Estimating Lagrange Multiplier}
This is the last step of the entire modeling (Fig. \ref{fig:steps}) and as new data comes in, this step will repeat. To compute the values for $\lambda_1 \text{ and } \lambda_2$, Limited-memory Broyden–Fletcher–Goldfarb–Shanno algorithm with Bounds (L-BFGS-B) is used mainly for two reasons. Firstly, it converges faster, secondly, and most importantly, it can natively enforce the strict bounds (Table \ref{tab:bounds}) unlike gradient descent \cite{cauchy1847methode}, Newton-Raphson \cite{raphson1702analysis} or other methods that are used for parameter estimation, especially in scenarios with highly incomplete or non-Gaussian data \cite{zhang2020maximum}. 


\begin{table*}[h!]
\centering
\caption{Details of the direct numerical bounds for the Lagrange multipliers ($\lambda_1 \text{ and } \lambda_2$) and lists the key physical and probability constraints that are enforced via penalties to ensure the algorithm finds a stable and realistic solution.}
\label{tab:bounds}
\begin{tabular}{c c c}
\toprule
\textbf{Type} & \textbf{Bounds} & \textbf{Implementation/Enforcement} \\
\midrule
Variable bounds & \begin{tabular}{@{}c@{}}$\lambda_1 \in (-\infty, -0.5)$, $\lambda_2 \in (0.1, \infty)$\end{tabular} & [($-\infty$ -0.5), (0.1, $\infty$)] \\
Physical constraints & $\lambda_1 + \lambda_2 UL_i < 0 \quad \forall i$ & Enforced via penalty \\
Probability bounds & $10^{-9} \le p^{Fij} \le 0.99$ & Enforced via penalty \\
\bottomrule
\end{tabular}
\end{table*}

The optimization objective is to minimize the following function:

\begin{equation*}
f(\lambda_1, \lambda_2) = \underbrace{(F_1 - PF)^2 + (F_2 - L)^2}_{\text{Target Function}} + \text{Total Penalty}
\end{equation*}

\noindent where:
\begin{equation*}
F_1 = \sum_{i,j} p_{ij} \cdot p^{Fij}, \quad F_2 = \sum_{i,j} UL_i \cdot p_{ij} \cdot p^{Fij}
\end{equation*}

\noindent and

\begin{equation*}
\begin{split}
&\text{Total Penalty} = \underbrace{\text{CONSTRAINT\_PENALTY} \cdot \sum (\lambda \text{ violations})^2}_{\lambda \text{ constraints}} \\
&\quad + \underbrace{\text{BOUND\_PENALTY} \cdot \sum (p^{Fij}\text{ bounds violation})^2}_{p^{Fij}\text{ constraints}}
\end{split}
\end{equation*}

The algorithm starts with a physics-informed guess for the lambdas to begin its search:.

\begin{equation*}
\lambda_1 = -1.5 \cdot \max(UL_i)\\
\text{, }\lambda_2 = 1.0
\end{equation*}

\noindent The optimizer calculates the gradient ($\nabla f$) to determine the steepest descent using the following approximation:

\begin{equation*}
\nabla f \approx \left[\frac{f(\lambda_1+\epsilon,\lambda_2)-f(\lambda_1,\lambda_2)}{\epsilon}, \frac{f(\lambda_1,\lambda_2+\epsilon)-f(\lambda_1,\lambda_2)}{\epsilon}\right]
\end{equation*}

\noindent where, $\epsilon$ is a very small number. Once $\nabla f$ is computed, the algorithm computes the Hessian matrix approximation,

\begin{equation*}
H_{k+1}^{-1} = (I-\rho_k \cdot s_k \cdot y_k^T)\cdot H_k^{-1}\cdot (I-\rho_k \cdot y_k \cdot s_k^T) + \rho_k \cdot s_k \cdot s_k^T
\end{equation*}
\noindent where,
\begin{itemize}
    \item  $H_{k+1}^{-1}$ refers to updated inverse Hessian. Approximation of the inverse curvature matrix at iteration k+1. Scales gradients for faster convergence.
    \item  $I$ denotes identity matrix. Matrix with 1s on the diagonal, 0s elsewhere. Base for matrix transformations.
    \item  $\rho_k$ is a scaling factor. $\rho_k = \frac{1}{y_k^T s_k}$. Ensures numerical stability in updates.
    \item  $s_k$ represents step vector. $s_k = x_{k+1} - x_k$. Captures parameter space movement.
    \item  $y_k$refers to gradient difference. $y_k = \nabla f(x_{k+1}) - \nabla f(x_k)$. Encodes curvature information.
    \item  $H_k^{-1}$ is previous inverse Hessian. Approximation from iteration k. Baseline for iterative improvement.
\end{itemize}
For the first iteration, the $y_k$ is $\nabla f$, $s_k$ is just the initial guess of lambda values and Hessian Matrix is the identity matrix.
It uses Hessian matrix to compute the direction:

\begin{equation*}
d_0 = -H_0^{-1} \nabla f_0
\end{equation*}
\noindent The algorithm then calculates the optimal bounded step size ($\alpha$) for updating:
\begin{align*}
\alpha =& \min(\frac{\lambda_1 \text{ (Updated Value) } - \lambda_1 \text{ (Previous Value) }}{|d_{\lambda_1}|}, \\&\frac{\lambda_2 \text{ (Updated Value) } - \lambda_2 \text{ (Previous Value) }}{|d_{\lambda_2}|})
\end{align*}

\noindent where, $d_{\lambda_1} \text{ and } d_{\lambda_2}$ are the search direction vectors for $\lambda_1 \text{ and } \lambda_2$ respectively. 

\noindent This step size is first used to update the Lagrange multiplier, $\lambda_1$:

\begin{equation*}
    \lambda_1 \text{ (Updated Value) } = \lambda_1 \text{ (Previous Value) } + \alpha \cdot d_{\lambda_1}
\end{equation*}

\noindent Similarly, the second Lagrange multiplier, $\lambda_2$, is updated in the iteration.

\begin{equation*}
    \lambda_2 \text{ (Updated Value) } = \lambda_2 \text{ (Previous Value) } + \alpha \cdot d_{\lambda_2}
\end{equation*}

The iterative process continues until one of the following convergence criteria is met:

\begin{itemize}
\item Function tolerance: $|f_{k+1} - f_k| < 10^{-3}$
\item Max number of iterations: 100
\end{itemize}

\section{Case Study and Results}

For this study, considered a scenario where an EV takes \textit{8 hours} to charge from 20\% to 80\%. To understand how much time was expected, we need to look at the vehicle and charging details. The car has a \textit{100 kWh} battery and was connected to the charger at \textit{20\%} and was disconnected from the charging station when the battery was \textit{80\%} charged. The energy needed (EN) for the session is calculated first using the Equation \ref{eq:energy}:

\begin{equation*}
EN = \frac{100 \text{ kWh}}{100} \times (80 - 20) = 60 \text{ kWh}
\end{equation*}

With the required energy known, the expected time can be computed using Equation \ref{eq:timetaken} based on the charging power of the level 2 charger, which was \textit{22 kW} and its efficiency, which was \textit{91\%}:

\begin{equation*}
\text{Charging Time (hours)} = \frac{60 \text{ kWh}}{22 \text{ kW} \times 0.91} \approx 3 \text{ hours}
\end{equation*}

The expected charging time is \textit{3 hours}. The difference between the actual time taken (\textit{8 hours}) and the expected time (\textit{3 hours}) confirms the 5 additional hours of time that defines \textit{the m=5} stress levels. To analyze this systemically, we consider a four-component system representative of the EV charging ecosystem: DSO (\textit{unit loss=18}), aggregator (\textit{unit loss=15}), CPO (\textit{unit loss=12}), and charging station (\textit{unit loss=9}) \cite{garofalaki2022electric}. We assume that the overall failure probability of the network is 0.45, and the total loss across all stress levels is 6 units. Additionally, we assume that the probability of a component experiencing stress (the probability of each component being responsible for the additional time) is given by Table \ref{tab:input}. \textbf{In this case study, the assumed values are chosen for demonstration only and are not from a specific dataset, a decision necessitated by the limited data availability in the literature. Nevertheless, these values were carefully constructed to reflect real-world data patterns.}

\begin{table}[ht]
\centering
\caption{$p_{ij}$ across components and stress levels}
\begin{tabular}{c c c c c}
\toprule
\textbf{Stress Level} & \textbf{DSO} & \textbf{Aggregator} & \textbf{CPO} & \textbf{Charging Station} \\
\midrule
0 & 0.066 & 0.082 & 0.091 & 0.056 \\
1 & 0.164 & 0.171 & 0.182 & 0.167 \\
2 & 0.230 & 0.223 & 0.227 & 0.231 \\
3 & 0.263 & 0.256 & 0.245 & 0.267 \\
4 & 0.277 & 0.268 & 0.255 & 0.279 \\
\bottomrule
\end{tabular}
\label{tab:input}
\end{table}

For these values and as described above, L-BFGS-B computes $\lambda_1 = -7.0859$ and $\lambda_2 = 3.9360 \times 10^{-1}$. The corresponding $p^{Fij}$ values are also computed and reported in Table \ref{tab:output}.


\begin{table}[ht]
\centering
\caption{$p^{Fij}$ across components and stress levels}
\begin{tabular}{c c c c c}
\toprule
\textbf{Stress Level} & \textbf{DSO} & \textbf{Aggregator} & \textbf{CPO} & \textbf{\begin{tabular}{@{}c@{}}Charging \\ Station\end{tabular}} \\
\midrule
0 & 0.367907 & 0.405318 & 0.456123 & 0.448627 \\
1 & 0.367948 & 0.450277 & 0.565533 & 0.664821 \\
2 & 0.367975 & 0.478819 & 0.628974 & 0.834057 \\
3 & 0.367989 & 0.497864 & 0.656300 & 0.947539 \\
4 & 0.367995 & 0.504975 & 0.671991 & 0.988699 \\
\bottomrule
\end{tabular}
\label{tab:output}
\end{table}

These resulting values fit the constraints of the model:
\begin{itemize}
    \item All $p_{ij}(\lambda_1 + \lambda_2 UL_i) > -1$: True (Equation \ref{eq:pFlimit})
    \item $\lambda_2 > 0$ True and $\lambda_1 < 0$: True (Equation \ref{eq:lambdavalues})
    \item $\lambda_1 < -\lambda_2 \times \max(UL_i)$: True (Equation \ref{eq:lambdavalues} and Equation \ref{eq:pFlimit})
    \item $p^{Fij}$ is increasing with $p_{ij}$ for all $UL_i$: True
\end{itemize}

\begin{figure}
    \centering
    \includegraphics[width=1\linewidth]{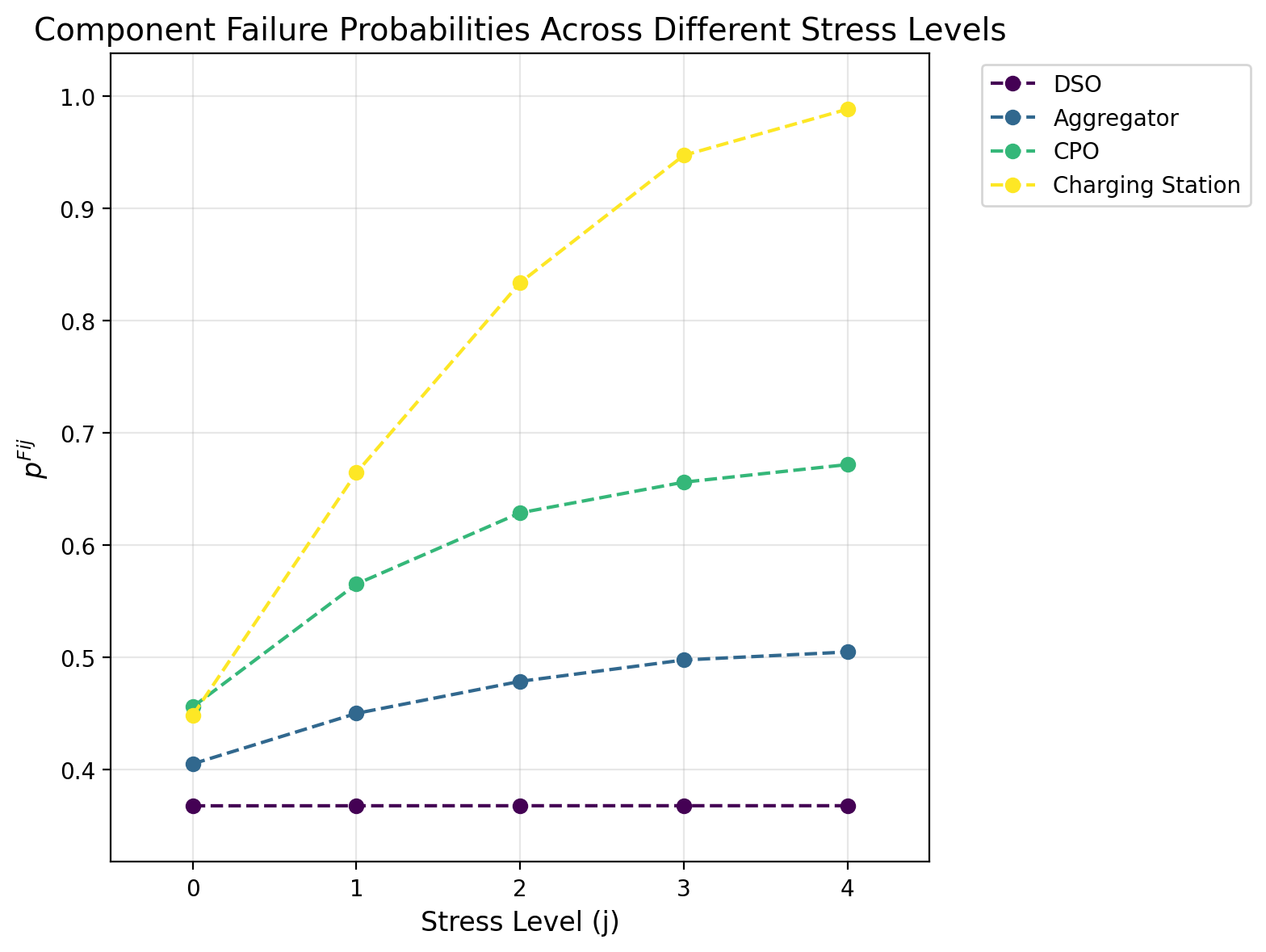}
    \caption{The graph illustrates the calculated failure probabilities for each of the four system components across increasing stress levels, highlighting the rapid approach to near certain failure in charging station.}
    \label{fig:pF}
\end{figure}

While Table \ref{tab:output} provides the precise failure probabilities for each component, visualizing this data allows for a more intuitive interpretation of the results. Fig. \ref{fig:pF} reveals patterns within our EV charging networks, validating the core principle of the our model, that failure probabilities ($p^{Fij}$) increase with stress levels but at dramatically different rates across components. The charging station is the most failure-prone component ($i^*$), reaching a failure probability of $0.9$ that is near-certain failure at maximum stress. This vulnerability also mirrors real-world scenarios where charging stations face the harshest operational conditions. Conversely, the aggregator and DSO demonstrate remarkable resilience, maintaining consistently low failure rates regardless of stress intensity as well as high $UL_i$. These asymmetric response patterns are mainly driven by varying unit losses ($UL_i$) and stress distributions ($p_{ij}$) as these are the only factors that differ between the components in this model.

\subsection{Analysis of Component Failure Probabilities}
\begin{figure*}
    \centering
    \includegraphics[width=1\textwidth]{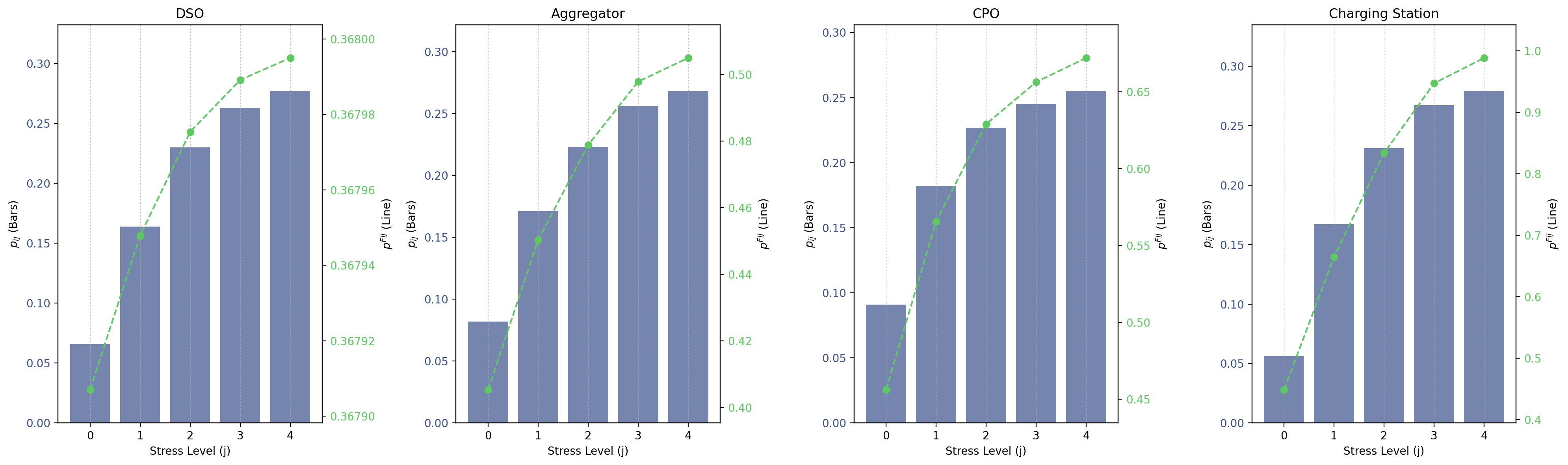}
    \caption{These plots detail individual component behavior, comparing the probability of experiencing a given stress level (bars) with the resulting failure probability (line), showing how $p^{Fij}$ increases with stress.}
    \label{fig:pvspF}
\end{figure*}

By examining the computed failure probability values, this analysis explores how distinct unit loss ($UL_i$) of each component and stress probability distribution ($p_ij$) drive its specific response to increasing operational stress. Fig. \ref{fig:pvspF} illustrates that the charging station, with a unit loss ($UL_i$) of 9, shows the steepest escalation in $p^{Fij}$, rapidly approaching 1 at higher stress levels. Conversely, the DSO, with the highest $UL_i$ of 18, exhibits the flattest curve, with $p^{Fij}$ remaining relatively stable across stress levels. The Aggregator ($UL_i=15$) and CPO ($UL_i=12$) show progressively steeper curves as their $UL_i$ decreases. This indicates that a component with a higher $UL_i$ can exhibit a faster rate of increase in $p^{Fij}$ even for small change in $p_{ij}$. This is demonstrated by Equation \ref{eq:pF} while not so prominent in Fig. \ref{fig:pvspF}, Table \ref{tab:input} and Table \ref{tab:output}, because we have assumed a lower values for $PF$. A higher $PF$ amplifies the rate of $p^{Fij}$ increase for critical components (with high $UL_i$) but has minimal impact on resilient components (with low $UL_i$) because high $PF$ will force $\lambda_1$ to be more negative forcing Equation \ref{eq:lambdavalues} to be more negative, and thereby, increasing the rate of change of $p^{Fij}$. The relationship between the stress probability ($p_{ij}$) of a component and its failure probability ($p^{Fij}$) is driven by two system-wide constraints: total network failure probability ($PF$) and total expected loss ($L$). These constraints are balanced by Lagrange multipliers ($\lambda_1$ and $\lambda_2$), which act as \textit{tuning knobs} adjusted during optimization. The speed at which $p^{Fij}$ rises with $p_{ij}$ depends on the term $-(\lambda_1 + \lambda_2 \cdot UL_i)$. This term is always positive (Equation \ref{eq:lambdavalues}), and a larger value means failure probability climbs faster under stress. More specifically:

\begin{itemize}
    \item \textbf{When both $PF$ and $L$ are low}, the system is stable and cost-effective. Here, constraints exert minimal pressure, so $\lambda_1$ is weakly negative and $\lambda_2$ is small. The term $-(\lambda_1 + \lambda_2 \cdot UL_i)$ remains modest, causing $p^{Fij}$ to increase slowly. Components tolerate minor stress spikes without rapid failure escalation since overall risk and costs are negligible.

    \item \textbf{When $PF$ is high but $L$ is low}, the system is unreliable but failures are inexpensive. To meet the high $PF$ constraint, $\lambda_1$ becomes strongly negative. This makes $-(\lambda_1 + \lambda_2 \cdot UL_i)$ large and positive, forcing $p^{Fij}$ to surge rapidly. Even small stress hikes trigger sharp failure probability jumps to meet the high failure points of the system, despite low financial impact.

    \item \textbf{When $PF$ is low but $L$ is high}, reliability is critical due to costly failures. $\lambda_1$ stays moderately negative, but $\lambda_2$ spikes positively to penalize expensive components. Consequently, $-(\lambda_1 + \lambda_2 \cdot UL_i)$ grows large, accelerating $p^{Fij}$ especially for high $UL_i$ components. Stress on critical parts must be minimized to avoid severe losses, even if failures are rare.

    \item \textbf{When both $PF$ and $L$ are high}, the system faces maximum pressure. $\lambda_1$ becomes strongly negative (to enforce high failures) and $\lambda_2$ rises, making $-(\lambda_1 + \lambda_2 \cdot UL_i)$ extremely large. $p^{Fij}$ escalates aggressively for all components. Any stress spike risks catastrophic failures and costs, creating a volatile, hypersensitive environment.
\end{itemize}

These delicate equilibriums sets the stage for examining system-level reliability analysis, where we evaluate how component interactions shape collective network behavior.

\subsection{System-Level Reliability Analysis}
\begin{figure}
    \centering
    \includegraphics[width=1\linewidth]{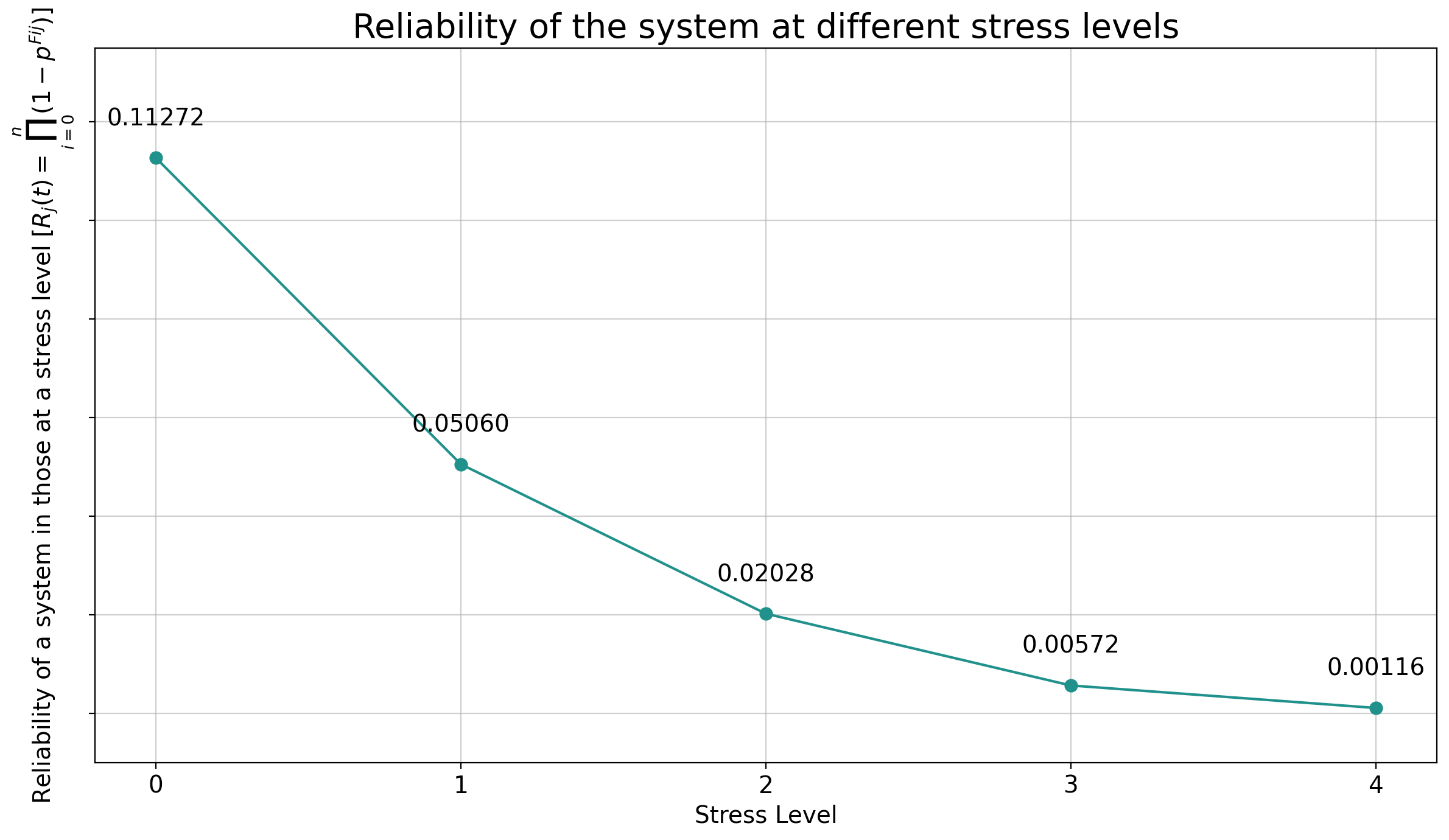}
    \caption{The graph demonstrates the reliability of the system at each discrete stress level (j), denoted as $R_j(t)$. This reliability decreases significantly as the stress level rises, dropping to nearly zero at the maximum stress. This steep decline is primarily driven by the failure probability of the charging station, which was reached $0.988699$ (refer Table \ref{tab:output}) at the final stress level.}
    \label{fig:R}
\end{figure}

Fig. \ref{fig:R} shows that the reliability $R(t)$ of an EV charging network decreases as component failure probabilities ($p^{Fij}$) increase, which is determined by Equation \ref{eq:reliability}. As $p^{Fij}$ rises due to additional time taken by an EV to charge the survival probability $(1 - \sum p_{ij} \cdot p^{Fij})$ for each component decreases. Since reliability $R(t)$ is the product of all component survival probabilities, even a single high $p^{Fij}$ value (e.g., near 1 for critical components like charging stations) disproportionately reduces $R(t)$.
Equation \ref{eq:pF} shows that higher stress probabilities ($p_{ij}$) and unit losses ($UL_i$) accelerate failure risks, causing $R(t)$ to approach zero rapidly. This inverse entropy-reliability relationship quantifies how rising unpredictability (entropy) degrades system's reliability.
\subsection{Key Findings and Insights}
Now that the experimental framework and mathematical relationships have been established, the insights from applying our PME based reliability model highlight some important patterns in system behavior. The interdependence of variables like $p^{Fij}, p_{ij} \text{ and } UL_i$ under entropy maximization, informs us how stress levels and unit losses dynamically shape failure probabilities. For instance, the constraint $\lambda_1 + \lambda_2 \cdot UL_i$ ensures that $p^{Fij}$ rises with stress levels \textit{(j)} reflecting real-world scenarios where prolonged operational delays (higher $p_{ij}$) amplify failure risks. Components that can have lower $UL_i$, such as the aggregator, exhibit flatter failure probability curves, while high-loss components like the charging point operator or power grid show sharp increases, nearing certainty of failure at extreme stress, as shown in Fig. \ref{fig:pF} and Fig. \ref{fig:pvspF}. This asymmetry shows how economic or operational penalties ($UL_i$) influence risk mitigation strategies, as the model penalizes deviations from aggregate constraints $PF$ and $L$ through Lagrange multipliers ($\lambda_1, \lambda_2$). These parameters, optimized via L-BFGS-B, act as balancing forces, aligning component-level failures with system-wide reliability targets. Since reliability is fundamentally a function of time, our model defines stress as additional time to serve as a generalized metric. While this is a fitting generalization, the predictive accuracy of the model could be enhanced by incorporating the specific, physics-based stressors that manifest as this additional time, such as the semiconductor degradation under high-power cycles analyzed by \cite{karunarathna2021reliability}.

The graphical analyses further contextualize these relationships, bridging abstract equations to tangible system performance. Fig. \ref{fig:pvspF} visually validates the core idea of our model that is rising stress levels correlate with higher $p^{Fij}$, but the rate of increase varies dramatically by component. charging station’s near certain failure at $j=3$ contrasts with DSO’s stability, illustrating how $UL_i$ and $p_{ij}$ distributions create component specific vulnerability profiles. Similarly, the reliability curve (Fig. \ref{fig:R}) quantifies the system’s reliability under stress, showing \textit{R(t)} decreases as stress increases. This inverse relationship between entropy and reliability also tells us about the trade-off inherent in complex systems, greater uncertainty reduces predictability, impacting overall robustness. The graphs highlight how even small stress increases at critical points can cause unexpectedly large drops in reliability, like a domino effect where minor disruptions spiral into system-wide failures. 

One of the key insight we see in our model is that the value for $p^{Fij}$ is restricted between $e^{-1}$ ($\approx$ 0.367) and $e^0$ ($\approx$ 1) under theoretically ideal conditions (Equation \ref{eq:lambdavalues} and Equation \ref{eq:pFlimit}). In the model implementation, constraints artificially cap $p^{Fij}$ between $10^{-9}$ and $0.99$. This gap reveals a subtle assumption of our model, that is the system always experiences some baseline stress, even at the lowest stress level ($j=0$), since $p^{Fij}$ cannot truly reach zero. While this reflects the reality of background operational risks in complex systems, it underscores the challenge of accurately modeling the stress effects in this framework. Similarly, the assumption in our model of a single-point failure is a fitting approximation since a single component fault can lead to total system failure and simultaneous failures occurring exactly at the same time are rare. Improving it to account for multi-component failures at the same time would still increase its versatility for a broader range of complex, real-world scenarios \cite{du2021maximum}.

The results confirm that the PME model works effectively for EV charging systems while remaining flexible enough for other uses. By taking total failure probability and expected loss as starting points, the method can handle varied failure scenarios and costs, making it useful beyond EVs. Its core idea is that using entropy to fill information gaps without assumptions can be applied broadly.

\section{Broader Applications and Implications} 
Though the methodology has been implemented within the context of the EV charging ecosystem, it is not limited to this application. The PME, along with reliability theory, offers a flexible framework that can be adapted to various systems dealing with partial information and uncertainty. By maximizing entropy under specified constraints, this approach can estimate probabilities of component failure or performance outcomes across a wide range of fields, making it a valuable tool for addressing reliability and optimization challenges in complex networks. Some of the possible applications are mentioned below.

\begin{itemize}
    \item \textbf{Smart Grid Energy Distribution: } The proposed methodology can be implemented in a smart grid system where different regions (components) experience varying levels of energy demand (stress). By knowing the overall reliability of the grid and the distribution of energy consumption across these regions, the methodology can estimate the probability of failures for different parts of the grid. This application would facilitate the optimization of energy allocation, allowing utility providers to predict potential overload scenarios and improve grid stability and efficiency.

    \item \textbf{Healthcare Resource Management: }The methodology can be extended to healthcare systems, particularly within hospital networks where departments (components) are subject to varying levels of patient inflow (stress). With knowledge of the overall performance metrics of the hospital and data on patient distribution, the model can estimate workload distribution across different departments. This predictive capability would enable hospital administrators to allocate resources more effectively, anticipate surges in patient demand, and enhance the overall quality of healthcare service delivery.

    \item \textbf{E-commerce Order Fulfillment: }The proposed approach can also be applied to e-commerce logistics networks, where fulfillment centers (components) handle different volumes of orders (stress). By utilizing data on the overall performance of the logistics network and the variability in order volumes across centers, the methodology can predict the likelihood of delays and inefficiencies at specific locations. This application would support strategic decisions in inventory management, route optimization, and overall supply chain reliability, leading to more efficient order fulfillment processes.
    
    \item \textbf{Urban Traffic Management: }The framework can analyze traffic density (stress) at intersections to predict congestion probabilities (failure). By treating road crossings as components and traffic flow as path sets, PME estimates traffic distribution patterns using known city-wide jam probabilities. This enables adaptive traffic light scheduling and route optimization to reduce bottlenecks, similar to how EV charging stress informs load balancing.

    \item \textbf{Agricultural Production Optimization: }PME can correlate environmental stressors (rainfall variability, fertilizer efficiency) with crop yield reliability. By partitioning agricultural land into sub-regions and treating crop failure as the ``network" collapse, the model estimates production distribution using average yield data and localized stress metrics. This mirrors the approach in the EV ecosystem to component-level stress analysis, helping farmers allocate resources to high-risk zones.

    \item \textbf{Network-Centric Defense Systems: }PME evaluates combat readiness under battlefield stress (enemy force strength, logistics delays). By modeling information-gathering processes and command nodes as components, the framework quantifies threat distributions using average combat support effectiveness. This parallels the EV cybersecurity analysis, where stress from cyberattacks informs reliability thresholds. Here, it could guide troop deployment strategies to protect critical network paths.

    \item \textbf{Financial Risk Assessment: }PME can adapt to model market volatility (stress) and portfolio failure probabilities. By treating economic indicators as constraints (analogous to EV charging station temperature limits), the framework estimates default risks using historical crash data and real-time stress signals like interest rate fluctuations. This entropy-driven approach avoids overfitting to past trends, just as EV models avoid assumptions about novel cyberthreats, enabling robust risk mitigation strategies.
\end{itemize}

The versatility of the methodology enables its adaptation to diverse scenarios, providing reliable and data-driven insights across multiple domains. Additionally, it highlights potential fields for future research that could further enhance the accuracy, robustness, and applicability of the model in an ever-evolving technological landscape.

\section{Conclusion and Future Work}
The PME-based reliability framework developed in this work addresses critical gaps in traditional EV charging ecosystem analysis by quantifying failure risks under data scarcity and system complexity. By maximizing entropy subject to known constraints, network failure probability (PF) and expected loss (L), the model generates unbiased failure probabilities ($p^{Fij}$) that reflect real-world dynamics. Unlike static indices like SAIDI/SAIFI, which lack adaptability, or Markov models, which ignore cyber-physical interdependencies, our approach integrates hardware degradation, software vulnerabilities, and grid interactions into a unified probabilistic framework. The inverse entropy-reliability relationship further allows the operators to quantify how unpredictability, whether from new cyberattacks or extreme weather or some other issue can degrade system robustness. By translating abstract mathematics into actionable metrics, such as optimized maintenance schedules or hardened infrastructure, this work bridges theoretical rigor with practical engineering, offering a scalable tool for evolving EV networks.


Despite these advancements, the model, like any focused analytical framework, inherits limitations inherent to its assumptions and scope. The framework makes the common assumption of single-point failures, where only one component can fail at a time. This standard approach simplifies the analysis but is not intended for the rare real-world scenarios where multiple components could fail at the exact same time simultaneously, which is a rare case. Real-world validation remains partial, a common challenge in the field that reflects the limited data availability in the literature. Although the case study was built to mimic real-world data and its simulations confirmed the strong theoretical consistency of our model, large-scale deployment across heterogeneous networks with protocols like OCPP and ISO 15118 is needed to assess generalizability. Since reliability is fundamentally a function of time, our model defines stress broadly as additional time to serve as a generalized metric. While it is a fitting generalization, robust framework of the model can be further improved by incorporating multiple forms of stress, such as the semiconductor degradation under high-power cycles. The model can also be readily updated to use deep learning methods to determine if there is a failure and if we need to implement the full model. These avenues for future work underscore the need for iterative refinements to align such a powerful framework with the different issues within the EV ecosystems.

Apart from improvement within the model, partnering with charging networks to deploy the framework on live infrastructure would validate its predictive power under diverse operational conditions, comparing model-derived $p^{Fij}$ values against actual downtime logs or component replacements. Also, incorporating repair kinetics would add more value to our work by providing probabilistic repair time as dynamic constraints, adjusting PF and L based on technician availability or spare parts logistics. Beyond EVs, adapting the methodology to smart grids could optimize transformer load balancing under renewable energy fluctuations, while healthcare applications might predict ICU equipment failures using patient inflow rates as stress analogs. Collaboration with insurers could also translate our reliability predictions into financial risk models. These extensions would not only enhance the relevance of the model to EV ecosystems but also establish it as a tool for reliability engineering in an increasingly interconnected world.

\section*{Acknowledgment}

The authors would like to extend their sincere gratitude to the Predictive Analytics and Technology Integration (PATENT) Laboratory and the Intellgent System and Predictive Analysis (ISPA) Laboratory for their support and resources throughout this research.


%





\ifCLASSOPTIONcaptionsoff
  \newpage
\fi





\bibliographystyle{IEEEtran}
\bibliography{IEEEabrv,Bibliography}

\vfill


\end{document}